\documentclass[10pt,journal,compsoc]{IEEEtran}

\usepackage{graphicx}
\usepackage[export]{adjustbox}
\usepackage{float}
\usepackage{subfig}
\usepackage{mathtools}
\usepackage{amsmath}
\usepackage{hhline}
\usepackage{stmaryrd}
\usepackage{nccmath}
\usepackage{tipa}
\usepackage{rotating}
\usepackage{tikz}
\usepackage{multirow}
\usepackage{tabularx}

\usepackage{color}
\usepackage{rotating}
\usepackage{siunitx}
\usepackage{multicol,blindtext}
\usepackage{lipsum}
\usepackage[bottom]{footmisc}

\begin{document}

\title{Exploring the Effectiveness of Service Decomposition in Fog Computing Architecture for the Internet of Things}

\author{Badraddin~Alturki,
        Stephan~Reiff-Marganiec,
        Charith~Perera,
        and~Suparna De 
\thanks{B. Alturki and S. Reiff-Marganiec are with the Department
of Informatics, University of Leicester, Leicester, UK e-mail:\{baba1, srm13\}@leicester.ac.uk}
\thanks{C. Perera is with the School of Computer Science and Informatics, Cardiff University, Queen's Buildings, 5 The Parade, Roath, Cardiff, CF24 3AA,  UK e-mail:charith.perera@ieee.org}
\thanks{S. De is with the Institute for Communication Systems, University of Surrey, UK,  UK e-mail:s.de@surrey.ac.uk }
}

\markboth{IEEE TRANSACTIONS Sustainable Computing, VOL. PP, NO. , PP PPPP}%
{Shell \MakeLowercase{\textit{Alturki et al.}}: Exploring the Effectiveness of Service Decomposition in Fog Computing Architecture for the Internet of Things}


\IEEEtitleabstractindextext{
\begin{abstract}
The Internet of Things (IoT) aims to connect everyday physical objects to the internet. These objects will produce a significant amount of data. The traditional cloud computing architecture aims to process data in the cloud. As a result, a significant amount of data needs to be communicated to the cloud. This creates a number of challenges, such as high communication latency between the devices and the cloud, increased energy consumption of devices during frequent data upload to the cloud, high bandwidth consumption, while making the network busy by sending the data continuously, and less privacy because of less control on the transmitted data to the server. Fog computing has been proposed to counter these weaknesses. Fog computing aims to process data at the edge and substantially eliminate the necessity of sending data to the cloud. However, combining the Service Oriented Architecture (SOA) with the fog computing architecture is still an open challenge. In this paper, we propose to decompose services to create \textit{linked-microservices} (LMS). \textit{Linked-microservices} are services that run on multiple nodes but closely linked to their linked-partners. \textit{Linked-microservices} allow distributing the computation across different computing nodes in the IoT architecture. Using four different types of architectures namely cloud, fog, hybrid and fog+cloud, we explore and demonstrate the effectiveness of service decomposition by applying four experiments to three different type of datasets. Evaluation of the four architectures shows that decomposing services into nodes reduce the data consumption over the network by 10\% - 70\%. Overall, these results indicate that the importance of decomposing services in the context of fog computing for enhancing the quality of service.
\end{abstract}


}
\maketitle

\section{Introduction}
\IEEEPARstart{A}{dvances} in the sensing and data processing capabilities of devices, coupled with that of communication networks is leading to the maturity of the Internet of Things (IoT) paradigm. A number of application domains, such as smart healthcare, smart homes and buildings, now rely on devices as varied as user smartphones, sensor network gateways and network routers for their realization. Most of these applications use the devices for sensing and data pre-processing tasks such as aggregation and filtering, with the majority of the data analysis done in centralized cloud infrastructures, which are located at the core of the communication infrastructure \cite{wang2017enorm}. With the predicted increase in the number of devices (28 billion connected devices by 2021 \cite{zhou2017sensors}, \cite{ericsson}), which can participate in existing and emergent applications, geographically centralised cloud data centers will find it difficult to support the highly distributed IoT devices without suffering a loss in QoS. The resultant massive flow and exchange of data from the large number of connected devices will also impact on electricity costs and carbon emissions \cite{kumar2018geintro}, with achieving energy efficiency a significant challenge. With typical IoT applications being highly context-dependent, the resultant short but high-frequency data communication pattern from participating devices will pose a challenge to the bandwidth of the communication and cloud frameworks  \cite{rahman2018semanticfog}. The existing computing and communication infrastructure is likely to cause unacceptably high latency in service delivery and network congestion \cite{Mahmud2017Latencyaware}, with recent studies showing that cloud servers geographically situated far from user devices affect latency more negatively than those geographically closer \cite{wang2017enorm}. IoT applications that collect sensitive data such as users' private information or location face the challenging decision of whether to store it locally or communicate it to the cloud, since securing the data will incur overheads and subsequently affect performance \cite{aazam2018}.

There has been a recent interest in moving away from centralized data processing centers to a more distributed fog computing paradigm to bring computing to the edge of the network, closer to user devices \cite{EdgeAnomaly, Hybrid}. Fog computing is defined as a hierarchically distributed computing paradigm that bridges cloud data centers and IoT devices \cite{bonomi2012fog}. The combination of IoT devices and fog computing enables smart environments that can respond to real-time events by combining services offered by multiple, heterogeneous devices. This can be achieved by decomposing the services into linked-microservices which can distribute the data processing as close as possible to the data source. Microservices are defined as independent, tiny autonomous services which function together to complete a task \cite{newman2015building}. It is worth noting that if the microservices are linked to each other, then the distribution of processes among fog nodes will be admissible in an IoT architecture. In other words, moving the computation from centralized approaches to more distributed ones will be possible, leading to a reduction in data communication cost over the network and reduced data frequency between fog devices and the cloud \cite{wang2017enorm, SmartGridReduction}. Most of such microservices can implement typical machine learning (ML) tasks that can deal with the volatility and heterogeneity of data produced in IoT environments. Data processing using ML techniques in typical IoT-fog applications consists of well-defined steps such as feature extraction, pre-processing and applying relevant algorithms. 

Though there have been advances made in fog computing with proposals for reference architectures \cite{openfog,mecarch2016}, practical realizations need to tackle the challenge of resource management \cite{wang2017enorm}. A recent study \cite{wang2017enorm} identifies that resource management at the edge of the network requires the following problems to be solved: provisioning fog nodes for executing workloads, managing computing, battery etc. resources on nodes and deploying workloads on nodes. All of these require knowledge and an awareness of the resources available on edge devices as well as constraints related to the services that run on such nodes. While automated service composition for Web services and for pervasive computing environments has been studied extensively \cite{cassar2013}, existing approaches do not focus on the costs related to time and reliability \cite{wang2017service}, while also ignoring service constraints or the data computation capabilities of the devices \cite{wang2014constraint}. Thus, there is a need for adapting the service composition approach for IoT domains by considering all the aspects (computation time, reliability and constraints). Moreover, the range of possible data computation capabilities in the IoT devices also needs to be taken into account when distributing service processes among the nodes.

Thus, in this paper, we focus on the research problem of arriving at the best service computation distribution strategy that is cognizant of node constraints and can deliver a reduction in data communication cost for different types of data modalities. For this, we conduct a range of experiments to see how the traditional machine learning algorithms perform in the fog computing domain to highlight the importance of efficiently selecting which services should run on which node. 

We acknowledge that theoretical analysis is important, however, we consider such analysis to be out of the scope of this paper. Additionally, our objective is to demonstrate the practical validation of the proposed approach. Our evaluation strategy is similar to the work that has been done by \cite{satyanarayanan2017emergence}.

We have used several machine learning algorithms including Naive Bayes, Logic Regression, K Nearest Neighbours (KNN), Decision Trees, Multi-layer perceptron (MLP) and Support Vector Machine (SVM) to explore and test which algorithm is the best fit for the given data modality. Based on the results, there is no single optimum technique for all types of datasets. Every algorithm has different training time, with some requiring less time or less storage such as Naive Bayes, KNN and Decision trees. However, in terms of executing the persistent features and multi-dimensional datasets, the SVM and MLP algorithms can perform them effectively. The authors in \cite{kotsiantis2007supervised} have reviewed many machine learning algorithms and they stated that it is not possible for an algorithm to perform better than others for all given datasets. This paper has surveyed the well known techniques with the focus being to find the key concepts. Therefore, our selection of machine learning algorithms was based on their efficiency in different datasets and being well known techniques.

\subsection{Sustainability}
Sustainability is important when deploying real-world systems because of several factors such as computation strategy, energy consumption, computation workload and data distribution strategy. The authors in \cite{perera2017fog} discussed ten crucial characteristics including data analytics, security and privacy, context awareness, mobility and other features to develop sustainable fog computing architectures. A sustainable system aims to optimise trade-offs when selecting the computation strategy, energy consumption and data communication usage. Thus, the proposed infrastructure can help develop sustainable computing architectures as it can enable handling of more computation workload at the network edge by distributing the data computation efficiently, which also has positive implications for energy consumption.

\subsection{Fog and Edge Computing}
Much of the literature mentions both fog computing and edge computing interchangeably \cite{chiang2016fog, aazam2018, habak2017}, with some papers stating that edge is a synonym of fog computing \cite{yi2015survey}. Both fog and edge paradigms agree on the concept of moving the computation as much possible from a centralized level to distributed or decentralized levels. However, authors in \cite{munir2017ifciot} stated that they differ in terms of the radio access network, with fog computing involving Wireless LAN (WLAN) or cellular networks, but edge computing is usually cellular. In this paper, we have used fog and edge computing interchangeably because of their similarity in moving the computation from centralised clouds to the edge of the network. The proposed linked-microservices decomposition strategy can be extended and applied to a range of edge devices, such as switches and routers if their device and data computation capability information can be obtained, as has been demonstrated in \cite{xu2018routercaching}.

\subsection{Contributions}
The range of fog nodes and data modalities (i.e. numerical, text and image) considered in our experiments are drawn from representative IoT-enabled fog computing applications such as crowd surveillance \cite{motlagh2017uav,motlagh2016survey}, service provision for massive ad-hoc crowds (e.g. 10 million Hajj pilgrims \cite{rahman2018semanticfog}), optimized computation distribution \cite{wang2017enorm}, augmented brain-computer interaction game \cite{ifogsim2017}, visitor-identification system in smart homes \cite{Mahmud2017Latencyaware} etc. The main contributions of this paper are as follows:

\begin{itemize}
\item We explore the importance of decomposing services into linked-microservices in a distributed architecture in fog computing domain for enhancing the QoS and meeting the low latency requirements of IoT-fog  applications.
\item We explore how different machine learning problems can be efficiently dealt with using service decomposition.
\item We propose an efficient approach, which decomposes the services and deploys them as close as possible to the edge of the network. We have conducted a trade-off analysis to demonstrate the usefulness of different service decomposition strategies, with the evaluation of the four possible strategies showing that decomposing service computation over the fog nodes reduces the data consumption over the network by 10\% (for text data) - 70\% (for numerical data).
This reduction in the data flow in turn implies less energy and bandwidth costs for the network, while also enabling reduced overheads for securing the condensed data features.

\item We have used different types of data modalities including numerical, text and images using different ML algorithms, since these are the most common modalities of IoT data sources, as shown from our analysis of a variety of IoT-fog  applications above. Thus, we believe that this is generalizable for most of the cases in this problem domain.
\end{itemize}

In this work our objective is to show that splitting the services or decomposing the services into microservices should  help in executing the services effectively, we propose as our future work to build a systematic way to divide the workload. For example, we tried different decomposition techniques and the results were different for each decomposition technique depending on the specific use case; for different kind of datasets the results will be different based on the results that we have achieved.

The remainder of this paper is organized as follows: section 2 reviews related works; section 3 presents a motivating scenario that is representative of the problem domain, while section 4 presents our methodology through the different datasets and ML algorithms considered for the experiments. Results are presented in section 5, followed by the corresponding evaluation and discussion in section 6. Pertinent research challenges are discussed in section 7, before we conclude the paper in section 8.

\section{Related Works}

\begin{table*}[!htbp]
\centering
\caption{Summary of Related Fog Computing Work}
\resizebox{18cm}{!}{
\begin{tabular}{|cccccc|} \cline{1-6}
\begin{tabular}[c]{@{}l@{}}
\centering

 \textbf{\huge Work}     
 \end{tabular}
&\begin{tabular}[c]{@{}l@{}}\multicolumn{1}{c}{\textbf{\huge Data}} \\ \multicolumn{1}{c}{\textbf{\huge Modality}} \end{tabular}
& 
\begin{tabular}[c]{@{}l@{}}\multicolumn{1}{c}{\textbf{\huge Fog Node}} \end{tabular}
& 
\begin{tabular}[c]{@{}l@{}}\multicolumn{1}{c}{\textbf{\huge Fog Node}} \\ \multicolumn{1}{c}{\textbf{\huge Functionality}} \end{tabular}  
&
\begin{tabular}[c]{@{}l@{}}{\textbf{\huge Distribution}} \\ \multicolumn{1}{c}{\textbf{\huge Strategy}} \end{tabular} 

& 
\begin{tabular}[c]{@{}l@{}}\multicolumn{1}{c}{\textbf{\huge Application}}\end{tabular}

\\ 
[2ex]
\hline

\multicolumn{1}{|c}{\textbf{\huge B. Tang, et al. \cite{tang2017incorporating}}}                                  
&       
\multicolumn{1}{c}{\textbf{\shortstack{\huge Real time Temperature \\ \huge sensing data}}}                                   
&                                                                 \multicolumn{1}{l}{\textbf{\shortstack{\huge Parallelized small  \\ \huge computing nodes}}}                                   
& 
\multicolumn{1}{c}{\textbf{\shortstack{\huge Data sensing,\\ \huge Pre-processing and \\ \huge data analysis}}}                                           
&
\multicolumn{1}{c}{\textbf{\shortstack{\huge The workload of \\ \huge Data analysis parallelized \\ \huge between edge nodes, \\ \huge parallel computing mechanism. }}}
&                                                                 \multicolumn{1}{c|}{\textbf{\shortstack{\huge Latency aware and \\ \huge location aware}}}                                                                                                                                        \\ [2ex]
\hline
\multicolumn{1}{|c}{\textbf{\huge P. Hu, et al. \cite{hu2017fog}}}  &      
\multicolumn{1}{l}{\textbf{\shortstack{\huge Three public face databases\\ \huge are used including \\ \huge Georgia Tech,\\ \huge Caltech and\\ \huge
BioID
}}}                                   
&                                                          
\multicolumn{1}{l}{\textbf{\shortstack{\huge Personal computer \\ \huge and Laptop }}}                                   
& 
\multicolumn{1}{l}{\textbf{\shortstack{\huge Face Detection,\\ \huge Pre-processing, \\ \huge data analysis and\\ \huge
computation offloaded \\ \huge from cloud
}}}                                     
&
\multicolumn{1}{c}{\textbf{\shortstack{\huge part of the computation\\ \huge  tasks is offloaded\\ \huge from cloud to\\ \huge fog nodes.}}}
&                                                                 \multicolumn{1}{l|}{\textbf{\shortstack{\huge Latency aware and\\ \huge network transmission \\ \huge sensitive}}}                                                                                                                                       \\ [2ex]
\hline
\multicolumn{1}{|c}{\textbf{\huge A. R. Zamani et al. \cite{zamani2017deadline}}}                                  
&       
\multicolumn{1}{c}{\textbf{\huge simulated data}}                                   
&                                                                 \multicolumn{1}{l}{\textbf{\shortstack{\huge Virtual Machines \\ \huge with resource limitations}}}                                 
& 
\multicolumn{1}{l}{\textbf{\shortstack{\huge Data Sampling \\ \huge and\\
\huge Camera Aggregator. 
}}  }                                           
&
\multicolumn{1}{l}{\textbf{\shortstack{\huge Resource federation model.\\ \huge Workload distribution \\ \huge based on value of data. \\ \huge Job Scheduling \\ \huge optimization strategy.}}}
&                                                                 \multicolumn{1}{l|}{\textbf{\shortstack{\huge Quality of Service,\\ \huge minimizing the computation\\ \huge cost and time\\ \huge of processing the job.\\ \huge  Reduce data transfer time.}}}                                          \\ [1ex]
\hline

\multicolumn{1}{|c}{\textbf{\huge Servia-Rodriguez et al. \cite{servia2017personal}}}          &       
\multicolumn{1}{c}{\textbf{\shortstack{\huge WISDM dataset \cite{kwapisz2011activity} \\ \huge (Numerical data), \\ \huge  Wikipedia \cite{WIKI2018} and NIPS \cite{NIPS2018}\\ \huge
 datasets (Text data)}} }                                  
 &                                                                 \multicolumn{1}{c}{\textbf{\shortstack{\huge Raspberry pi and\\ \huge  Personal devices}} }        
 & 
\multicolumn{1}{c}{\textbf{\shortstack{\huge Supervised and \\ \huge Unsupervised Learning.}}}                                           
&
\multicolumn{1}{c}{\textbf{\shortstack{\huge Share model (Training model)\\ \huge is distributed from\\ \huge cloud to personal devices\\ \huge  to create personal model }}}
&                                                                 \multicolumn{1}{c|}{\textbf{\shortstack{\huge Privacy preserving,\\ \huge improving accuracy\\ \huge and \\ \huge maintaining Efficiency.}}}                                                                                        \\ [2ex]
\hline
\multicolumn{1}{|c}{\textbf{\huge Ni, Lina, et al. \cite{ni2017resource}}}                                    &       
\multicolumn{1}{c}{\textbf{\huge Simulated data}}                                   &                                                                 \multicolumn{1}{c}{\textbf{ \huge Linux based Computers}}                                   & 
\multicolumn{1}{c}{\textbf{\shortstack{\huge Fog can select \\ \huge satisfying resource from \\ \huge previously allocated \\ \huge resources}}}                                             &
\multicolumn{1}{c}{\textbf{\shortstack{\huge resource allocation\\ \huge strategy  based on\\ \huge Priced Timed Petri Nets \\ \huge (PTPN)}}} 
&                                                                 \multicolumn{1}{c|}{\textbf{\shortstack{\huge Aware of task \\ \huge completion \\ \huge price and time.}}}                                                                                                                                         \\ [2ex]

\hline
\multicolumn{1}{|c}{\textbf{\huge N. Wang, et al. \cite{wang2017enorm} }}                                  &                                               
\multicolumn{1}{c}{\textbf{\shortstack{\huge Open dataset \\ \huge and iPokeMon game \cite{pokemon}.}}   }                                &                                                                 \multicolumn{1}{c}{\textbf{\shortstack{\huge Resource Allocation,\\ \huge request managing,\\ \huge communication latency \\ \huge monitoring,\\ \huge allocate or deallocate\\ \huge resources to containers.}}}                                   & 
\multicolumn{1}{c}{\textbf{\shortstack{\huge Linux based Containers}}}                                             &
\multicolumn{1}{c}{\textbf{\shortstack{\huge Offloading workloads \\ \huge from cloud to fog.}}}
&                                                                 \multicolumn{1}{c|}{\textbf{\shortstack{\huge Latency Aware}}}                                                                                                                                        \\ [2ex]
\hline
\multicolumn{1}{|c}{\textbf{\huge R. Mahmud et al. \cite{Mahmud2017Latencyaware} }}                                  &                                               
\multicolumn{1}{c}{\textbf{\shortstack{\huge Simulated data}}}                                   &                                                                 \multicolumn{1}{c}{\textbf{\shortstack{\huge Heterogenous nodes in\\ \huge modelled environment}}   }                                & 
\multicolumn{1}{c}{\textbf{\shortstack{\huge Intermediate layer\\ \huge between cloud \\ \huge and IoT devices}}}                                             &
\multicolumn{1}{c}{\textbf{\shortstack{\huge Optimization of resources }}}
&                                                                 \multicolumn{1}{c|}{\textbf{\shortstack{\huge Latency Aware}}}                                                                                                                                        \\ [2ex]
\hline

\multicolumn{1}{|c}{\textbf{\huge D. G. Roy et al. \cite{roy2017application} }}                                  &                                             
\multicolumn{1}{c}{\textbf{\shortstack{\huge Experimental data}}}                                   &                                                                 \multicolumn{1}{c}{\textbf{\shortstack{\huge Laptop, Personal Computer \\ \huge and mobiles.}}   }                                & 
\multicolumn{1}{c}{\textbf{\shortstack{\huge Cloudlet agent,\\ \huge offloading tasks \\ \huge to mobile phones.}}}                                             &
\multicolumn{1}{c}{\textbf{\shortstack{\huge Offloading applications \\ \huge from cloud to \\ \huge multi-cloudlet. }}}
&                                                                 \multicolumn{1}{c|}{\textbf{\shortstack{\huge Latency Aware and \\ \huge power consumption \\ \huge is reduced}}}                                                                                                                                        \\ [2ex]
\hline

\end{tabular}

}

\caption*{}
\end{table*}

This section reviews the current state of the art from the two aspects relevant to the problem definition: a. various computation distribution strategies employed in fog computing and the associated architecture implementations and b. service decomposition as has been researched in the wider Web services paradigm. 

The architectural aspect related to the location of the data processing is important. Data processing can be applied in various architectures including centralized, decentralized and distributed as follows \cite{castanedo2013review}:

\begin{itemize}
\item  In a centralized architecture, all sensors' data will be transmitted to the cloud for processing. It is widely accepted that the cloud has unlimited processing power which allows processing large data in an effective manner. Nevertheless, in a real-time case study, the data communication over the network will be huge, which will lead the cloud to be insufficient for efficient data fusion. In addition, this architecture will be more tricky when dealing with image data. The reason is that the data arrival time will be delayed which will adversely affect the result.

\item  In a decentralized architecture, there are different nodes with diverse computational capabilities in the network, accordingly, there is no central server similar to the centralised system. A node can apply data fusion autonomously to the local data and the data that are obtained from peers or neighbours. One of the major disadvantages of decentralized architecture is the huge cost of communication among peers. In this regard, there might be lack of scalability when increasing the number of nodes.

\item In a distributed architecture, sensors data will be processed at the source prior applying data fusion to a particular node which has the capability to fuse the data. This can cope with different problems of the centralised architecture and reduce the high cost of communication among peers in the decentralized architecture. However, this architecture can bring several challenges such as data distribution which needs to meet the flexible requirements under situations that are not expected. In addition, privacy and security can be one of the challenges that distributed architecture can face as the data will be saved in different locations, but security and privacy can be considered as a challenge that most of the architectures face \cite{PrivacyEngineering}. 
\end{itemize}



In recent years, there has been an increase in the amount of literature on distributed architecture in the IoT. One of the attempted proposals is by \cite{tang2017incorporating}; they proposed distributed architecture for fog computing for analysing big data in a smart city. They distribute the smartness to the devices in edge and computation at every layer which executes applications that have latency awareness. A fog computing based face resolution framework is proposed in \cite{hu2017fog} which obtains the information by analysing facial image. There are several features of this framework, including reduced data communication over the network and the response time of resolution, also efficiently solving the issues with bandwidth.  A proposal in the distributed analysis by \cite{zamani2017deadline} which is a model that combines processing power which is at network level including edge and data centres to process and analyse the data from collection point to a destination. Furthermore, in \cite{servia2017personal}, personal modal training method has been proposed in which the data processing, particularly machine learning is applied to private data in devices that have constraints and raspberry pi was used to test the feasibility of the IoT device in the implementation of such methods. Authors in \cite{wang2017enorm} proposed a framework for managing edge nodes which is called Edge NODE Resource Management (ENORM). In addition, they proposed several techniques that provision edge node resources. They used online game called PokeMon Go-like to check the feasibility of their framework. Their results show that by using ENROM the application latency is reduced between 20 - 80\%. In \cite{Mahmud2017Latencyaware} authors proposed an approach (latency aware) to place application modules on fog nodes to make sure that the service delivery satisfies the deadline for diverse applications. They modelled and evaluated their policy in Fog environment that is simulated in iFogSim \cite{ifogsim2017}. In resource allocation for fog computing, authors in \cite{ni2017resource} proposed effective resource allocation approach depending on Priced Timed Petri Nets (PTPN) for fog computing influenced by online shopping sales. Furthermore, the users can select the required resources dynamically from already allocated resources. Also, they showed an algorithm that predicts the cost of time and price for finishing jobs relies on PTPN structure.

Table 1 shows a summary of the related works of fog computing, which are reviewed along the following aspects: 
\begin{itemize}
\item Data Modality: format and modality of the data being used for implementation and validation.
\item Fog Node: types of fog nodes that are considered (user phones, low power embedded devices, general purpose computers, high-performance computers etc.) 
\item Fog node functionality: functions that the fog nodes perform such as data sensing and pre-processing tasks, computation offloaded from cloud.
\item Distribution Strategy: strategies used to distribute services, workload or computation to fog nodes from other nodes or from the cloud. 
\item Application: context aspects  considered in the computation/service distribution.
\end{itemize}
 It can be seen that several data modalities and formats are used, ranging over numerical, text and image data. Different types of devices are used as fog devices to perform different functions. The distribution strategy used is typically tied to the application scope and focus, ranging from data analysis parallelization, computation offloading to different optimization strategies.



 In web service composition, there is one proposal that focuses on autonomous web service composition by taking care of service constraints \cite{wang2014constraint}. It is important to be aware of the constraints on services, but in the IoT, there are nodes which also have various constraints. This means that there is a need to have a constraint awareness approach for both services and nodes. In addition, \cite{stelmach2013service} investigated the service composition's requirements and the way to obtain a composite service using transport domain as an example. They provided several scenario based approaches to service composition and discussed these. Additionally, authors in \cite{chen2011comprehensive} proposed a comprehensive device collaboration model which has four layers, namely device, device-oriented web service (doWS), resource and process. This model shows the possibility of integration between devices and web services, also the devices can be considered as active actor because the data is not sent immediately to servers. The authors in \cite{wang2015discovering} proposed a service based model in requirements decomposition, their model process starts with user requirements which are defined as goals, service discovery and discovered services employed to check the feasibility of the preceding decomposition. They decomposed the requirements of three web service composition cases to validate their approach. Another proposal in service decomposition domain by \cite{athanasopoulos2015cohesion} who proposed a greedy algorithm that decomposes interface into interfaces depending on cohesion. This approach mainly focused on improving the cohesion and it was successful in that regards, but other aspects are not considered like coupling between interfaces.     

Furthermore, authors in \cite{khanouche2016energy} proposed quality of service aware and energy centred service selection algorithm for service composition in the Internet of Things. The idea of the algorithm that it is possible to save energy by decreasing the degree of quality of service while maintaining the expectation of the user. The algorithm has two phases, the first one performs pre-selecting the services by proposing quality of service degree which meets the user's needs. In the second one, in order to select the best option among selected services, a relative dominance relation has been used for the process of service composition. Additionally, \cite{rodriguez2012data} proposed a method to perform data fusion via service composition model of DOHA (Dynamic Open Home-Automation) which is SOA-based middleware in a distributed manner. Every service is liable to get data from outer services using composite processes to control, fuse or create new information. In the implementation, they used DPWS (Device Profile Web Service) which is a framework that develops lightweight service for constrained devices. This framework put restrictions on web service specifications that allow the web services to run on resource-constrained devices, for example, the size of messages. In \cite{akasiadis2015developing} authors investigated the possibility of building complex services in IoT environment. They showed the SYNAISTHISI IoT adaptable platform that is able to combine services, devices and people with systems. The services in this platform are enriched semantically using ontologies. They developed intelligent meeting room ontology and presented the way that the developer can create a service, that defines how many people are inside an intelligent room. Authors in \cite{shadija2017microservices} examined the problems of microservices granularity and how it affects the latency. They simulated the deployment of microservices with two approaches including microservices in one container and microservices divided into several containers. They observed a slight increase in latency for several containers over single container deployments.


 Our review of the existing literature shows that most of the proposed models, approaches and architectures do not take into account resource constraints of the IoT devices. We acknowledge that many works have been done on data processing in resources constraint devices, although none of them have proposed service decomposition as a viable solution. The proposals in the service computing domain usually do not focus on deployment of services on nodes by considering the constraints of devices; instead mostly focusing on the quality of services and constraint requirements of services. However, in the case of IoT systems, there will be many constrained devices distributed across the network. This means it is important to decompose the resource-heavy services into smaller micro-linked services which can be distributed to and handled by constrained devices.


\section{Problem Analysis}
IoT devices have constraints on resources like RAM, CPU and storage and the services that execute on these devices have restrictions expressed in terms of the same resources. Additionally, service execution and distribution also needs to take into account the data computation capabilities (i.e. in terms of the installed library support) of the devices. Therefore, we need to model the data about services to get the knowledge about their restrictions before distributing them across the nodes. For example, the image recognition service needs at least 500MB to be executed, so the IoT device’s capability should be powerful enough to execute this service. Therefore, if the device has a limitation in processing power, then it is not possible to execute the service on it. In this case, service decomposition is an important aspect of the IoT architecture due to the involvement of devices with limited hardware capabilities and varying data computation availability, which cannot handle resource intensive tasks. In addition, decomposing services into linked-microservices is an important aspect in terms of service composition in the IoT architecture. This is crucial for effective distribution of services to the nodes in the IoT. The main challenge is to determine which services should be executed on which node in a given IoT architecture, by considering both overall efficiency and feasibility. This is similar to Job shop problem which is one of the most known problems in combinatorial  problems \cite{graham1966bounds}. Basically, the idea of the 'job shop' problem, is that there are a group of machines with varying levels of computational power (i.e., given a specific job, j, it could be the case that there are two machines that complete j in different amounts of time, with the quicker one having higher computational power). The problem then asks for an algorithm which produces an optimal assignment of jobs to machines, such that the overall amount of time it takes for all jobs to be completed is minimal.  The following scenario illustrates the problem by using a real use case. The scenario is drawn from a recent UAV (unmanned aerial vehicle) crowd surveillance study \cite{motlagh2017uav}, which looked at energy efficiency achieved by offloading the facial recognition operation to a Mobile Edge-Computing node rather than processing it locally on the UAV (using a Raspberry-Pi for computation). We extend this in our case to include multiple fog nodes (devices) with varying hardware and computation abilities. The scenario demonstrates the significance of deploying the right service on the right node.
 \begin{figure}[!b]
  \centering
   \begin{minipage}[b]{0.48\textwidth}
    \includegraphics[width = 8.8cm, height = 5.5cm, frame]{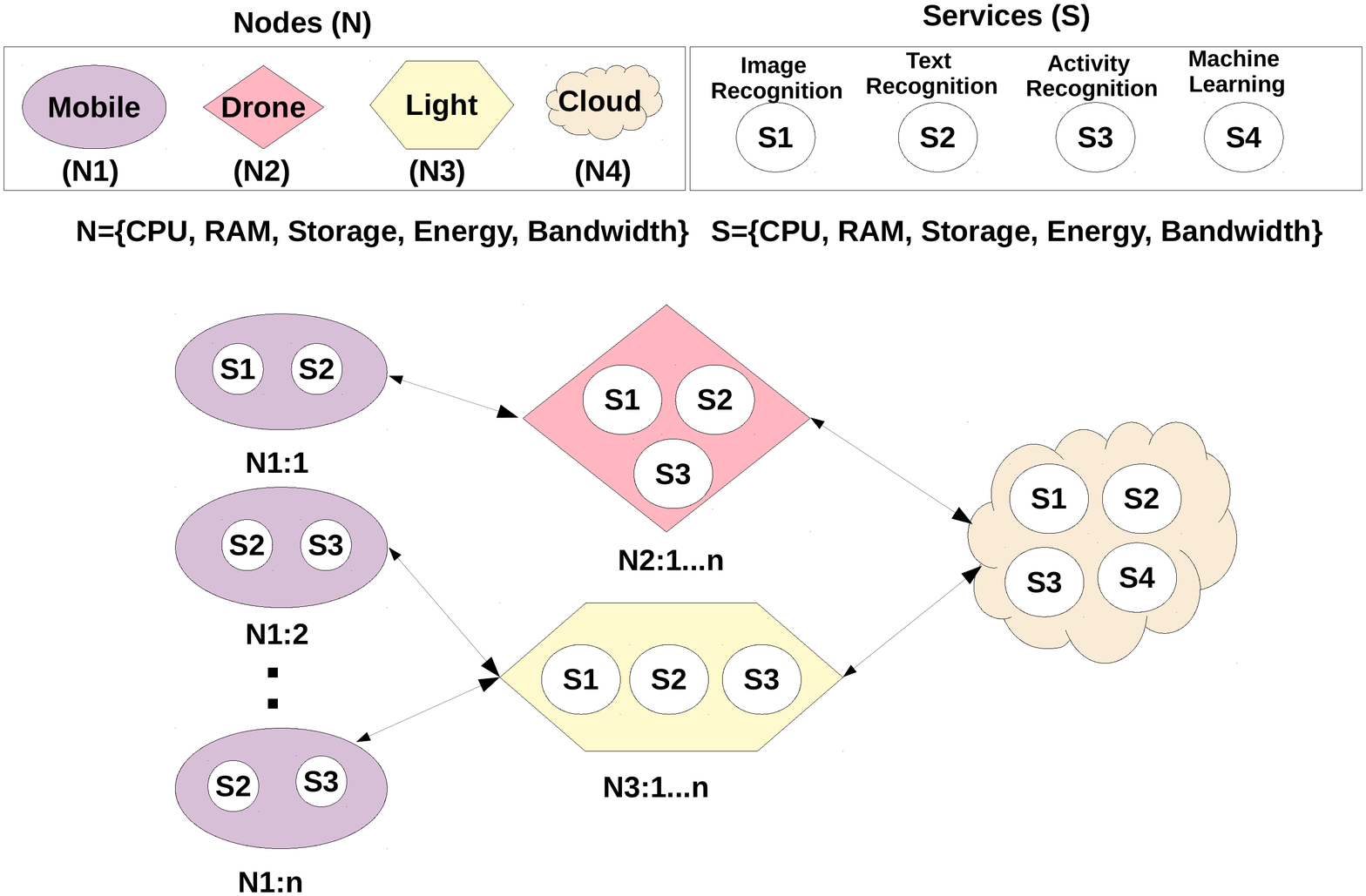}
  \end{minipage}
\caption{Scenario: City and Particular Event}
\label{scenario}
\end{figure}


\subsection{Motivating Scenario: City and Particular Event}
There is a major event in a city where people are taking videos and photos. The law enforcement agencies are interested in re-utilizing the captured images to identify criminals (or person of interest) among the crowds in order to anticipate crimes.
In  Figure \ref{scenario}, there are four types of Nodes ($N_i$) including mobile phones ($N_1$), drones ($N_2$), street lights ($N_3$) and cloud ($N_4$). Every node has a different combination of resources (CPU, RAM, energy, storage and network bandwidth) and each node can execute several different services ($S_i$). Each service requires a specific combination of resources to be executed on a given node. Additionally, the facial recognition service comprises a variety of linked-microservices corresponding to ML tasks, including facial feature extraction, data fusion, data filtering and face detection algorithms. The event goers are taking the photos and videos of the event primarily for their own pleasure, not for helping the law enforcement agencies. However, they may like to help the law enforcement agencies to maintain safety as long as the primary function of their devices in this scenario is not compromised.

The sustainability comes from better energy consumption, less communication means longer duration and more devices can be connected together.
The overall architecture should not consume too much energy or communication bandwidth from users' devices.  The simplest case of service distribution will send all the raw data to the cloud from individual user devices even though it consumes a lot of network bandwidth. However, if we can do the data transformation in a smartphone, then it consumes less bandwidth and sends only processed data to the cloud. The services associated to the face recognition ML tasks are deployed dynamically in smartphones. It is crucial to consider which responsibilities should be assigned to smartphones. In addition, the nearest lamp or drone will have more computing power than the phones and can handle the more computation intensive tasks than user devices. This reduces the communication cost in two ways: first, the images will be transformed into feature vectors in the mobile, and secondly, further processing will be applied to data when they are sent to the lamps or drones which are temporarily deployed because of the event and connected with the mobiles via Bluetooth which is cheap in terms of communication. Then the transformed data will be sent to the cloud via 3G for further analysis. 

This scenario introduces a number of challenges since it involves deploying dynamically composed services during an event in the city. The event happens on a particular day and people are likely to move around while taking photos which introduces unpredictability in their location. The service orchestrator needs to be aware of the resources available on the users’ phones when sending a request for service computation to them. It is crucial to consider how to compose services like sending data to the drones as mediator. Additionally, what services should be deployed to the drone and mobile is also an important aspect. This scenario illustrates the problems involved in service provisioning on fog nodes (taking into account varying hardware and data processing capabilities) and deploying the service taking into account data communication costs.

\begin{table}
\centering
\caption{Parameters}
\begin{tabular}{|l|l|} 
\hline
\textbf{Parameter~} & \textbf{Description}                                                                                                         \\ 
\hline                                                                             
$s_i$, $s_i$ $\in$ $\{$ $s_1$, $s_2$, $s_n$ $\}$   & Services  \\
\hline
$N_i$, $N_i$ $\in$ $\{$ $N_1$, $N_2$, $N_n$ $\}$   & Nodes  \\
\hline
\end{tabular}
\end{table}
\section{Methodology}
As we discussed in the literature, there are clear trade-offs among the three architectures (centralized, decentralized and distributed). It is possible to find solutions that can maximize the advantages and minimize the disadvantages of each architecture. Our proposed method endeavour to fulfill this, which is presented below. We propose an efficient approach which aims to move the computation from the cloud to the fog as much as possible. 

\begin{figure}[!b]
  \centering
   \begin{minipage}[b]{0.48\textwidth}
    \includegraphics[width = 8.8cm, height = 5.65cm]{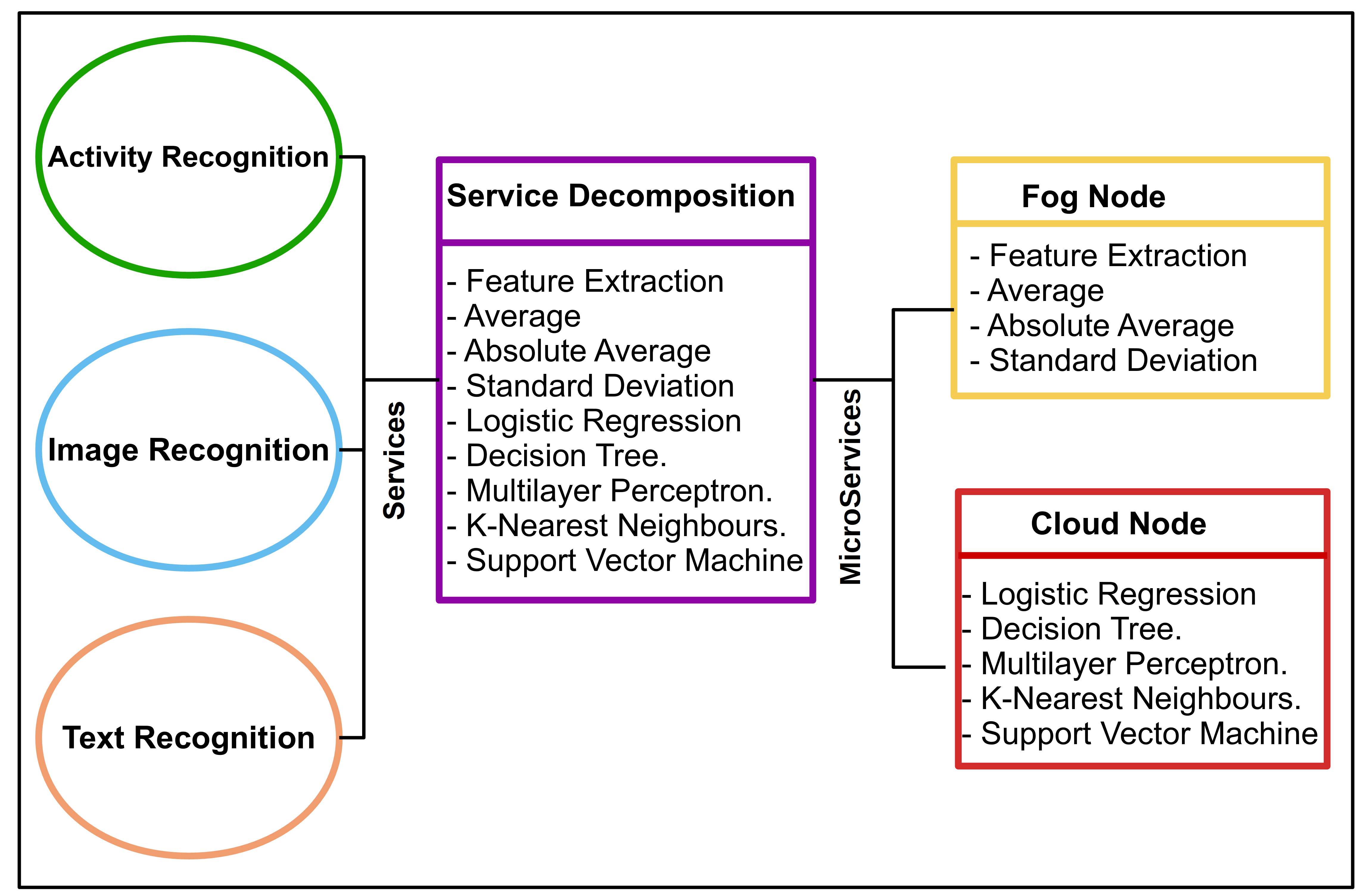}
  \end{minipage}
\caption{Service Decomposition}
\label{Decomposition}
\end{figure}

The goal G is to process a service $s$ in an effective and efficient manner.
\\
\\
\centerline{  $s_i$ $\shortrightarrow$ G($s_i$)}
\\
\\
We begin with decomposing services into set of linked-microservices MS to distribute them among nodes in our architecture. 
\\
\\
\centerline{ $\forall$ $s_i$, $s_i$ $\shortrightarrow$ MS($s_i$)}
\\
\\
A good illustration of this is shown in Figure \ref{Decomposition}, there are three services namely Activity $s_1$, image $s_2$ and text $s_3$ recognition. These services will be decomposed into linked-microservices ($MS$) before the distribution process. 
Then, the distribution process will distribute the micro services depending on the constraints on services and nodes of the nodes. Then we apply data processing techniques to sensors' data to extract features and reduce the number of data points in the fog node ($FN$). 
\\
\\
\centerline{ $MS$($s_i$)  $\shortrightarrow$  FN($MS$($s_i$)) }
\\
\\
This phase is significant because it is not possible to analyze raw time-series data by algorithms of classification effectively. Then, the cloud will receive the extracted features for creating inferences and training purposes. 
\\
\\
\centerline{ $MS$($s_i$)  $\shortrightarrow$  CN($MS$($s_i$)) }
\\
\\
Therefore, to get the results we compose both the fog node and the cloud node to process the service.
\\
\\
\centerline{FN($MS$($s_i$)) , CN($MS$($s_i$)) $\shortrightarrow$ G($s_i$)}
\\
\\
We conducted experiments for each of the above architectures in order to explore how the hybrid data analytics architecture would be beneficial in a variety of ways.

We have used three types of datasets namely numerical, text and image. The details of each dataset including dataset description, algorithms and process will be discussed below. Each dataset has its own description and algorithms. However, they have a common process in terms of decomposing the services and distributing the computation over the nodes. It is worth discussing the similarities before discussing the details of each experiment.

 \begin{figure*}[htb]
  \centering
  \includegraphics[width=18cm, height=8.8cm]{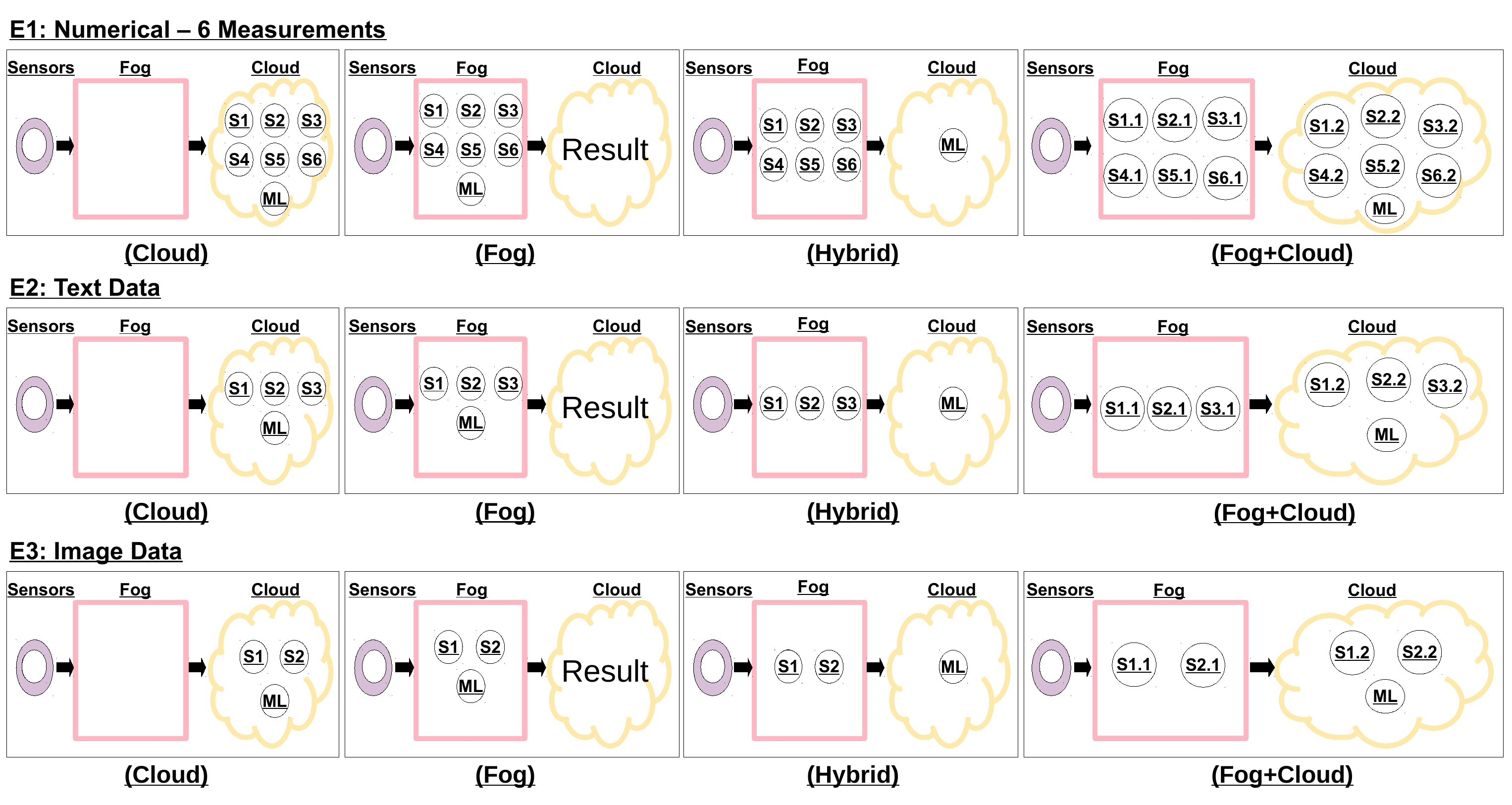}
 \DeclareGraphicsExtensions.
 \small{S = Statistical Measurements as Services, ML = Machine Learning Algorithms as Services}
 \\
  \small{(Cloud) = all the services are processed in the cloud, (Fog) = all the services are processed in the fog, (Hybrid) = the statistical measurement and machine learning services are distributed into fog  and cloud respectively, (Fog+Cloud) = 70\% of statistical measurements services processed in the fog and the rest with machine learning services are processed in the cloud. }
\caption{Process of Three Experiments. }
\label{processofthree}
\end{figure*}

\subsection {Process of three types of experiments}
We conducted 4 sets of experiments under each of the three types of experiments which are  Numerical, Text and Image data as shown in Figure. \ref{processofthree}.
\paragraph*{First Experiment} we sent all raw data to the cloud and data transformation methods are applied on the raw data to extract features. Then, we applied Machine learning to  the altered data based on the extracted features in the cloud as shown in Figure. \ref{processofthree} (\{1, 2, 3\}: \{Numerical, Text Data, Image Data\}. (Cloud)). We calculate the accuracy of each algorithm, the amount of data that is sent to the cloud and the execution time. 

\paragraph*{Second Experiment} we applied data transformation methods to the raw data to create features. Then, we applied analytical algorithms  to the modified data based on the extracted features in the fog as shown in Figure. \ref{processofthree} (\{1, 2, 3\}: \{Numerical, Text Data, Image Data\}. (Fog)). We check how feasible is the resource constraint device when processing the data and we measure the time of the execution.  

\paragraph*{Third Experiment} we applied the feature extraction methods to the raw data to extract features in the fog. By applying this the data is minimized as much possible in the fog, then the transformed data will be sent to the cloud for further analysis. as shown in Figure. \ref{processofthree} (\{1, 2, 3\}: \{Numerical, Text Data, Image Data\}. (Hybrid)). We calculate the accuracy of each algorithm, the amount of data that is sent to the cloud and the time of execution.

\paragraph*{Fourth Experiment} it is similar to the third experiment in terms of applying data aggregation algorithms to the raw data in order to extract features in the fog. However, in this approach, we applied the feature extraction on part of the raw data in fog and the remaining in the cloud. We divided the dataset into two parts randomly, where 70\% of the data will be used in fog and the remaining 30\% will be sent to the cloud. In this experiment, the statistical measurements in the Fog are presented as S1.1, S2.1, S3.1, S4.1, S5.1 and S6.1, which are applied on the 70\% of the raw data. However, in the Cloud the statistical measurements are presented as S1.2, S2.2, S3.2, S4.2, S5.2 and S6.2, which are applied on 30\% of the raw data. In numerical data, in the first part, we applied fusion methods on the 70\% of the raw data which is 768746 rows that is equal to 35 MB (70\% of the file size) in the fog. Then, we sent the transformed data which is equal to 0.84 MB and remaining raw data (30\% of the data) which is equal to 15 MB to the cloud for data transformation and then analysis as shown in Figure. \ref{processofthree} (E1: Numerical - 6 Measurements. (Fog+Cloud)). In text data, the first part has 14 newsgroups which are 70\% of the data and it will be processed in the fog. The second part has 6 newsgroups which are 30\% of the data and it will be transferred to the cloud. Then, we sent the transformed data and remaining raw data to the cloud for further data transformation and analysis as shown in Figure. \ref{processofthree} (E2: Text Data. (Fog+Cloud)). In image data, the first part is training data which have nearly 70\% of all images and equals to 17185 images (392 MB file size) and it will be processed in the fog. The second part is testing data which have nearly 30\% of all images and equals to 7815 images (178 MB file size) and it will be sent to the cloud. Then, we sent the transformed data and remaining raw data to the cloud for further analysis as shown in Figure. \ref{processofthree} (E3: Image Data. (Fog+Cloud)). In this experiment, we calculate the accuracy of each algorithm, the amount of data that is sent to the cloud and the time of execution.

\subsection{Experiment 1: Numerical Data with 6 measurements}
\subsubsection{Dataset Description}
The dataset is called WISDM \cite{kwapisz2011activity} which is accelerometer data that are gathered from  volunteers (36 users) who are performing six activities (jogging, walking, descending downstairs, climbing upstairs, sitting and standing). The volunteers held their mobile phones (Android based) when they were doing the six activities for a period of time. The collected data is divided into 10-second chunks. Additionally, 43 features are extracted relying on every 200 readings, in which every reading contains three acceleration values (x, y and z), in the fixed chunks. The transformed data contain 5418 accelerometer traces. In addition, the average traces per volunteer is around 150 and the standard deviation is around 44.
\subsubsection{Algorithms}
We have used six statistical measurements that are used in \cite{kwapisz2011activity} as shown in Figure \ref{processofthree}: E1: Numerical - 6 Measurements. There are 43 features that are created including the mean (S1), standard deviation (S2), average absolute difference (S3), time between peaks (S4) of every axis , average resultant acceleration of all axis (S5) and binned distribution for each axis (10 equal sized bins and total 30 bins) (S6). After preprocessing the data, five methods of classification (ML) are applied including Naive Bayesian (NB), Logistic Regression (LR), K-Nearest Neighbours (KNN), Decision Tree (DT) and Multilayer Perceptron (MP).


\subsection{Experiment 2: Text Data}
\subsubsection{Dataset Description}
The dataset is called Twenty Newsgroups \cite{Lang95} which is a set of nearly 20,000 newsgroup files. This dataset is collected for a different purpose, but it has become common data for text analysis in machine learning environment. In addition, the data are divided into 20 newsgroups and each group has a different topic.  
\subsubsection{Algorithms}
To apply analytical algorithms to text data, it is important to convert the text into a numerical feature vector. To extract features, first, we will use Tokenizing text with scikit-learn (S1) which has text pre-processing and filtering. Second, from occurrences to frequencies (S2) which counts the occurrences and divide the available occurrences of every word by the whole words of the file. These features have a specific name which is term frequencies (tf). Also, downscaling (tf-idf: Term Frequency time inverse document frequency) the weights for words that appear in most of the files have less knowledgeable information than the words that appear in very small part of the file.    
After extracting features from data, it is possible to apply the classifier (ML) to give a prediction of the category in the post. The first classifier is naive Bayes classifier in scikit-learn which has a variety of the classifiers and the most suitable is a multinomial variant for this dataset. The second classifier is support vector machine (SVM) which can be considered one of the most used algorithms in text classification.

\subsection{Experiment 3: Image Data}
\subsubsection{Dataset Description}
Dogs vs. Cats dataset \footnote{https://www.kaggle.com/c/dogs-vs-cats} which was a competition from Kaggle, is used and the purpose is to have the classification of cat and dog, so we can know if the picture has a dog or cat. The dataset is divided into two parts including training and testing data. The total number of images in the dataset is 25000 which is equal to 570 MB. 

\subsubsection{Algorithms}
To apply machine learning algorithms, we need to convert the images into a feature vector. We are going to use two methods that take input and produces feature vector as output. First, the $image\_to\_feature\_vector$ (S1) function that takes the image as input and changes the size of the image to stable height and width and the intensity level of RGB is converted into a single set of numerical data. Second, $extract\_color\_histogram$ (S2) function gets an image as input and produces the histogram of colour to describe the image colour classification. Then, we have used k-nearest neighbours algorithm (k-NN) (ML) classifier to give a prediction of the category as either dog or cat.

While doing all the three experiments, our expectation was that we can achieve a similar accuracy with the hybrid approach while maintaining the processing time and with considerably minimising data communication over the network.

\begin{figure*}[!htbp]
\centering
\begin{minipage}[b]{.22\textwidth}
\centering{\includegraphics[width=4.2cm, height=2.3cm, frame]{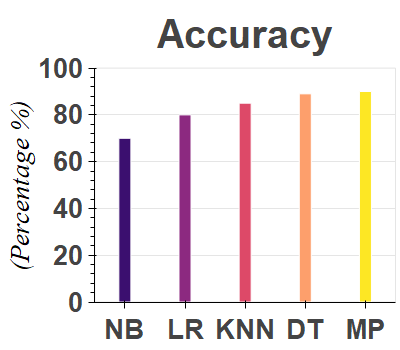}}
\caption*{(a)}
\end{minipage}\qquad
\begin{minipage}[b]{.22\textwidth}
\centering{\includegraphics[width=4.2cm, height=2.3cm, frame]{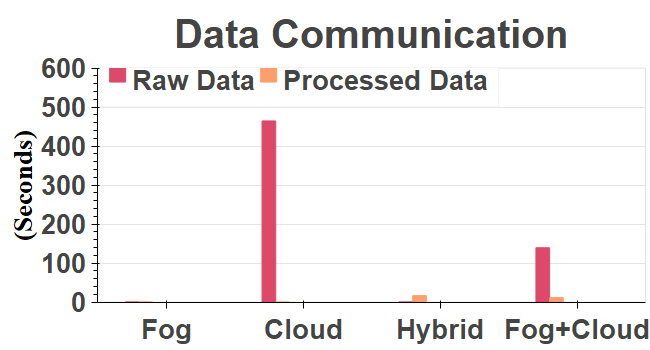}}
\caption*{(b)}
\end{minipage}\qquad
\begin{minipage}[b]{.22\textwidth}
\centering{\includegraphics[width=4.2cm, height=2.3cm, frame]{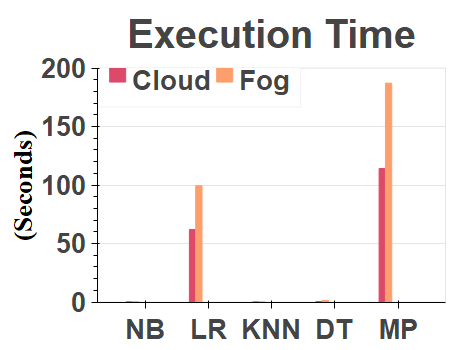}}
\caption*{(c)}
\end{minipage}\qquad
\begin{minipage}[b]{.22\textwidth}
\centering{\includegraphics[width=4.2cm, height=2.3cm, frame]{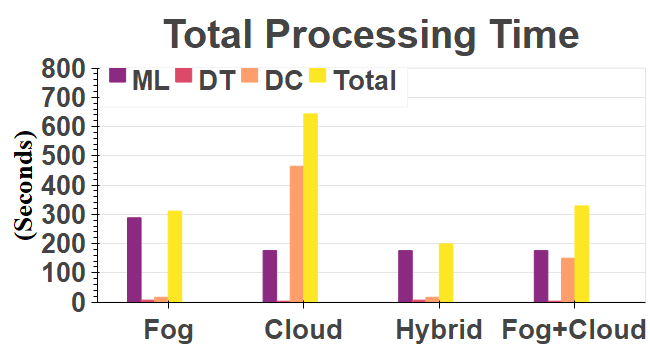}}
\caption*{(d)}
\end{minipage}
\small{NB = Naive Bayes, LR = Logistic Regression, KNN = K Nearest Neighbours, DT = Decision Tree J48, MP = Multilayer Perceptron}
\caption{Experiment 1: Numerical - 6 Measurements}
\label{Experiment 1}
\end{figure*}

\begin{figure*}[!htbp]
\centering
\begin{minipage}[b]{.22\textwidth}
\centering{\includegraphics[width=4.2cm, height=2.3cm, frame]{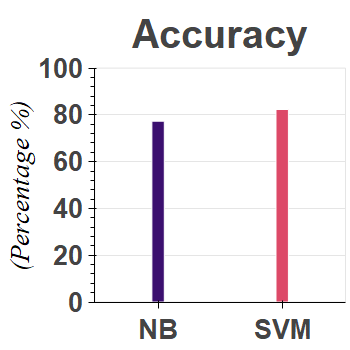}}
\caption*{(a)}
\end{minipage}\qquad
\begin{minipage}[b]{.22\textwidth}
\centering{\includegraphics[width=4.2cm, height=2.3cm, frame]{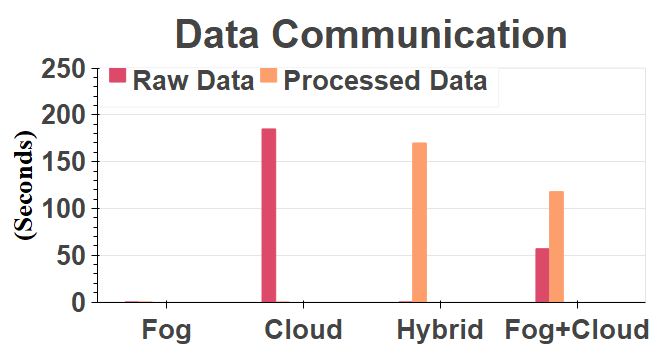}}
\caption*{(b)}
\end{minipage}\qquad
\begin{minipage}[b]{.22\textwidth}
\centering{\includegraphics[width=4.2cm, height=2.3cm, frame]{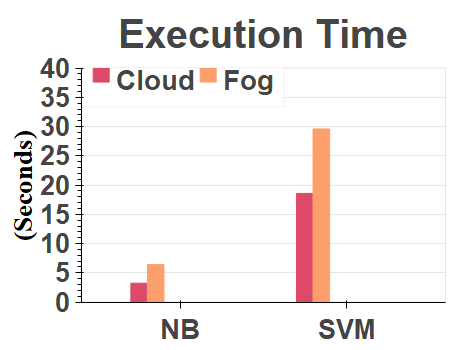}}
\caption*{(c)}
\end{minipage}\qquad
\begin{minipage}[b]{.22\textwidth}
\centering{\includegraphics[width=4.2cm, height=2.3cm, frame]{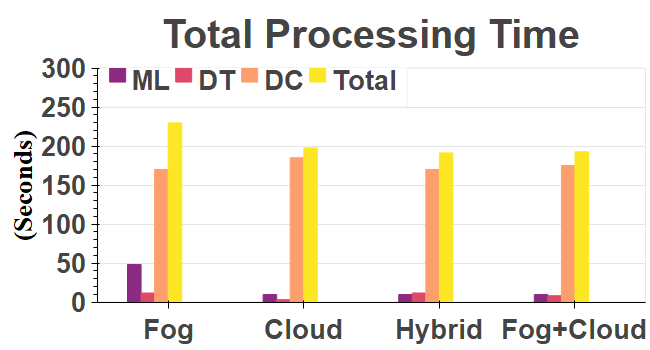}}
\caption*{(d)}
\end{minipage}
\small{SVM = Support Vector Machine, NB = Naive Bayes}
\caption{Experiment 2: Text Data}
\label{Experiment 2}
\end{figure*}

\begin{figure*}[!htbp]
\centering
\begin{minipage}[b]{.22\textwidth}
\centering{\includegraphics[width=4.2cm, height=2.3cm, frame]{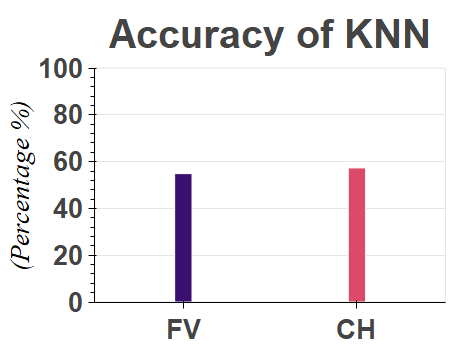}}
\caption*{(a)}
\end{minipage}\qquad
\begin{minipage}[b]{.22\textwidth}
\centering{\includegraphics[width=4.2cm, height=2.3cm, frame]{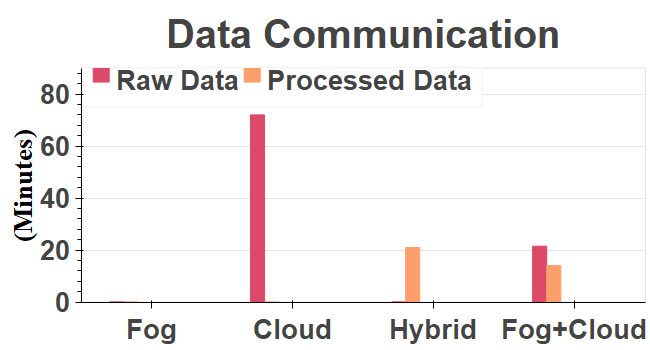}}
\caption*{(b)}
\end{minipage}\qquad
\begin{minipage}[b]{.22\textwidth}
\centering{\includegraphics[width=4.2cm, height=2.3cm, frame]{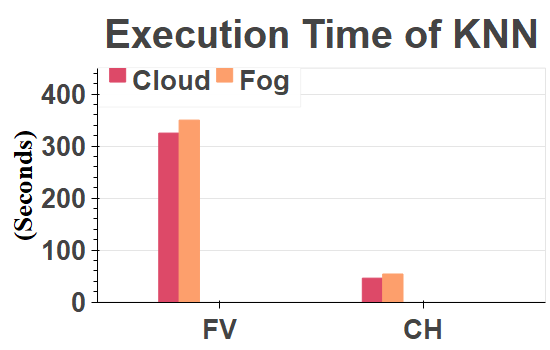}}
\caption*{(c)}
\end{minipage}\qquad
\begin{minipage}[b]{.22\textwidth}
\centering{\includegraphics[width=4.2cm, height=2.3cm, frame]{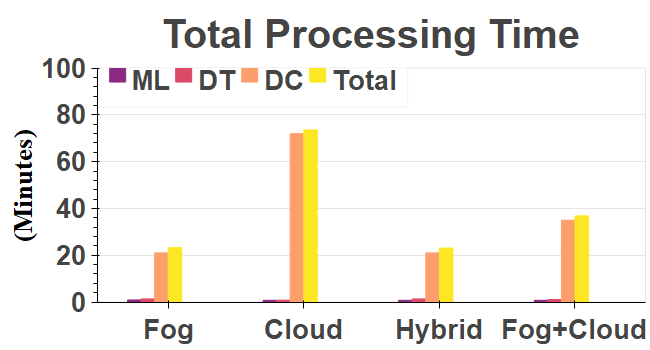}}
\caption*{(d)}
\end{minipage}
\small{KNN = K Nearest Neighbours, FV = Image to Feature Vector, CH = Extract Color Vector }

\caption*{ML = Machine Learning, DT = Data Transformation, DC = Data Communication}

\caption*{Summary: $(a)$ the accuracy of the analysis algorithms that are applied to the transformed data.
$(b)$ the time of data communication between the fog and cloud.
$(c)$ the time of execution of the analytical algorithms in both devices the fog and cloud.
$(d)$ demonstrates the total processing time for the four architectures.

}
\caption{Experiment 3: Image Data}
\label{Experiment 3}
\end{figure*}

\section{Experimental Set up and Results}

As discussed earlier, creating features from raw data such as numerical data, text data and image data, helps in the process of applying analytical algorithms on them. In our experiments, we have used a Raspberry Pi 3 model as the resource-constrained device. The device specification is 1GB RAM, with Raspbian Jessie as the operating system. These experiments can be performed on smartphones as well, but in our experiments we preferred Raspberry pi as it has similar specifications to smartphones, and is much cheaper that smartphones, with the cost also being a factor in IoT environments. Some papers have obtained and process data on smartphones, as in \cite{servia2017personal, wang2017enorm, nanxichen2018}. Furthermore, a Linux based System which has 16GB RAM, is used to mimic the device of the cloud. For data aggregation, segmentation and feature extraction, we have used Java and python libraries. 

We used python 2.7.12 and 3.5.2 and the weka 3.8 tool for machine learning (classification methods). For the numerical experiments, the weka tool is used and the heap size is adjusted in both cloud and fog. In the cloud environment, the size of heap was set to 8GB, whereas in the fog, the size was 650MB RAM for the numerical data experiment. However, the remaining experiments were done in python. While conducting these experiments, the internet upload speed was 1 Mbps. We used several packages and libraries in python for image analysis including Numpy, argparse, OpenCV packages, scikit-learn library and imutils library. In addition, scikit-learn library and Numpy packages are used for text analysis. 

\subsection{Experiment 1: Numerical - 6 Measurements}

   The results show that the highest percentage of accuracy achieved is in multilayer perception. In addition, both algorithms logistic regression and multilayer perceptron have important diversity between the two sides. Clearly, in the execution time, the cloud takes less time than the fog to perform ML algorithms because of its unlimited power. The Total Processing Time graph has three calculations for every architecture including the ML algorithms' execution time, the data transformation's execution time and the time of data communication between the cloud and fog as shown in Figure \ref{Experiment 1}. $Experiment$ 1: Numerical - 6 Measurements. $(d)$. This graph's results have been summarized in Table 3.

\subsection{Experiment 2: Text Data}
The results show that the highest accuracy is obtained by Support vector machine. There are two bars visible in Figure \ref{Experiment 2}. $Experiment$ 2: Text Data $(b)$: one for the transformed data and the other for the raw data. It is obvious that in the fog only device, the processing happens locally, therefore there is no data communication cost over the network. However, in terms of data communication from fog to cloud, the cloud approach was the highest as all raw data are sent. On the other hand, in the hybrid approach, the data communication from fog to the cloud is lower than both cloud and fog+cloud approaches as only the transformed data are transmitted. This result emphasizes that it is possible to reduce data communications by pre-processing the data early in the network. 
 The results show us that the two algorithms (Support Vector Machine and Naive Bayes) have considerable diversity between the two sides. It is clear that the cloud takes less time than the fog to execute ML algorithms because of its unlimited power. The Total Processing Time graph has three calculations for every architecture including the ML algorithms' execution time, the data transformation's execution time and the time of data communication between the cloud and fog as shown in Figure \ref{Experiment 2}. $Experiment$ 2: Text Data $(d)$. This graph's results have been summarized in Table 3.

\subsection{Experiment 3: Image Data}
 The results show that applying K-NN classifier on \textit{extract\_color\_histogram} has higher accuracy percentage than \textit{Image\_to\_feature\_vector}. There are two bars visible in Figure \ref{Experiment 3}. $Experiment$ 3: Image Data. $(b)$: one for the transformed data and the other for the raw data. It is obvious that in the fog only device, the processing happens locally, therefore there is no data communication cost over the network. However, the cloud approach was the highest in terms of  data communication from fog to cloud as all raw data are transmitted. On the other hand, the hybrid approach was lower than both cloud and fog+cloud approaches in terms of data communication from fog to the cloud as only transformed data are transmitted . This result emphasizes that it is possible to reduce data communications by pre-processing the data earlier in the network. It is clear from the results that applying K-NN classifier \textit{Image\_to\_feature\_vector} significantly takes more time to execute than \textit{extract\_color\_histogram}. However, comparing between cloud and fog there is no big difference in execution as in data communication. Obviously, the cloud takes less time than the fog to execute classification algorithms due to  its unlimited processing power as shown in Figure \ref{Experiment 3}. $Experiment$ 3: Image Data. $(c)$. The Total Processing Time graph has three calculations for every architecture including the ML algorithms' execution time, the data transformation's execution time and the time of data communication between the cloud and fog as shown in Figure \ref{Experiment 3}. $Experiment$ 3: Image Data. $(d)$. This graph's results have been summarized in Table 3.

  \begin{figure*}[!htbp]
\centering
\caption*{TABLE 3: Total Processing Time of the Three Experiments}

\subfloat{
    
     \includegraphics[width=16cm,height=9cm]{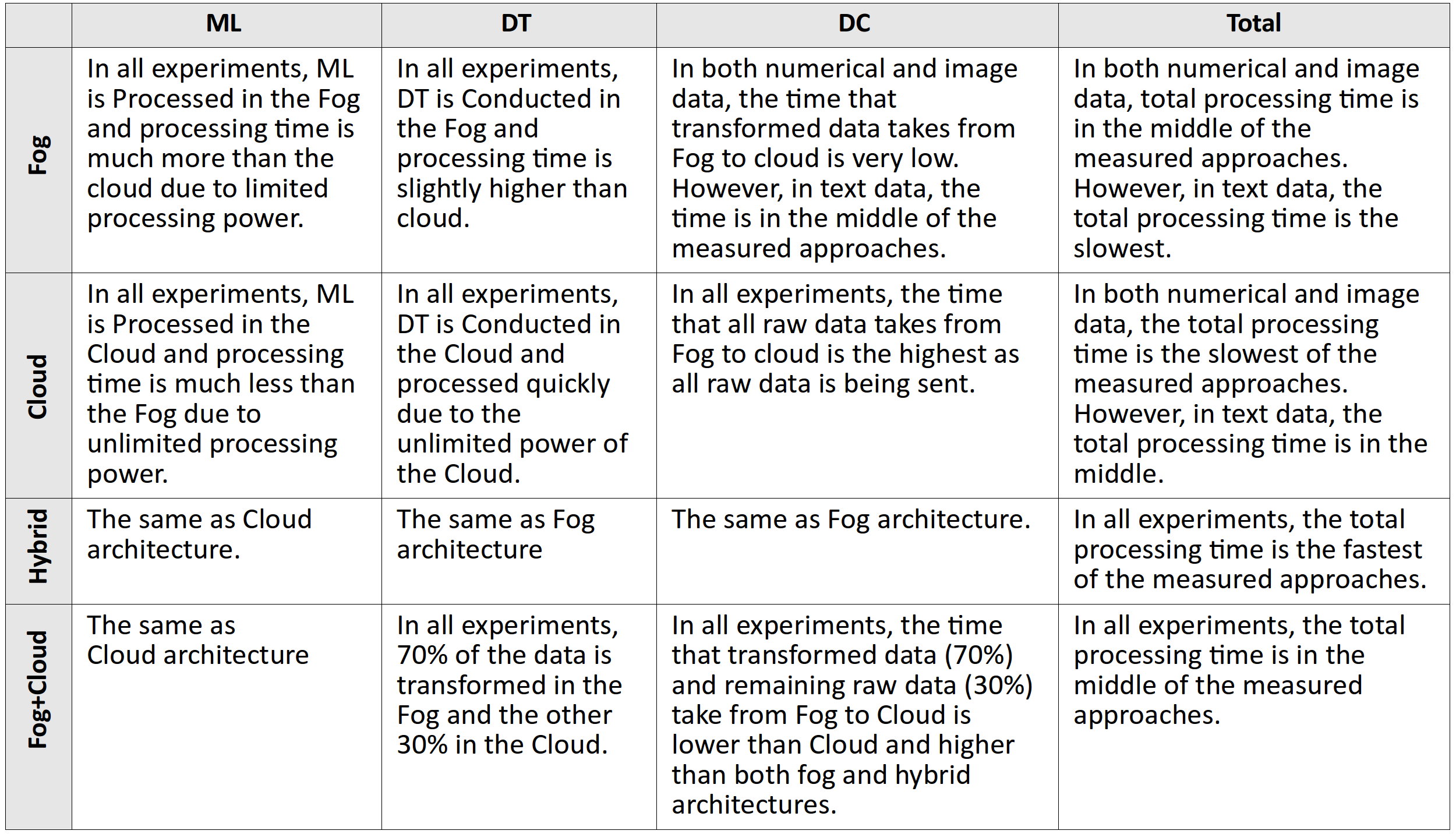}
 }
 \caption*{ML = Machine Learning, DT = Data Transformation, DC = Data Communication}
\end{figure*}
\section{Evaluation and Discussion}
The three approaches Fog, Hybrid and Fog+Cloud approaches have similarities in most cases. whereas, the difference between them is the location of the processing machine learning algorithm. In the Hybrid approach, the data transformation processing is done in the fog node, and then the transformed data are transmitted from fog to cloud to apply classification algorithms to the processed data. However, in the fog approach, the data transformation and machine learning process have been done at the fog node itself. The hybrid approach has an advantage over the fog approach that is utilizing the cloud's processing power for applying more sophisticated algorithms that require further processing power such as machine learning algorithms. Thus, this phase is important for minimising the processing time which has an effect on the total processing time. The results present that the proposed efficient approach is perfect for the given datasets and algorithms. As it can be seen that the results show data communication over the network is effective and gives considerable gains.

\textbf{The first observation: Data communication over the network.} It is widely accepted that when we increase the data size, the data communication over the network will be higher and costly. 
\paragraph*{Experiment 1: Numerical Data} In this experiment, there were approximately 1 million rows of raw data that are collected from mobile phones which nearly equal to 50MB in terms of data size. On the contrary, after we applied data fusion algorithms on the raw data we extracted features, and the rows of data are reduced to 5418 rows which equals 1.2 MB data size.
\paragraph*{Experiment 2: Text Data} In this experiment the raw data were around 20000 newsgroups file which equal to 22.4 MB. After extracting features, the size is decreased but not significantly when we used the method to convert text into a numerical feature vector. Maybe we do not have large savings, but we created features which will allow us to do more analysis. 
\paragraph*{Experiment 3: Image Data} In this experiment, the raw data were around 570 MB. However, after extracting features from images, the size became of 170 MB. This shows that extracting the features from images in network level can help with significant savings.

According to the experiments, significant savings can be achieved in data communication over the network by applying data transformation earlier in the network. This observation will be more important when the quality and number of sensors rise significantly and therefore the resolution and rate of data will be growing quickly. By applying data fusion to the data in the fog/edge node near to the data source before they are sent to the cloud, we will reduce the data and send only meaningful data. By doing this, the energy consumption in fog devices will be reduced, these devices obtain their internet connection through networks like 3G, 4G or even 5G, therefore the devices' batteries will be lasting longer too.

\textbf{The second observation: Accuracy of each algorithm.} In the results, transformed data can be seen as less accurate than working with the raw data. 
\paragraph*{Experiment 1: Numerical }Overall, the accuracy loss is between 7 - 25\% which is not excessive: the lowest level of accuracy is 75\%, however, the highest level accuracy is approximately 93\%. 
\paragraph*{Experiment 2: Text Data} Overall, the accuracy loss is between 17 - 25\% which is not extreme: the lowest level of accuracy is 77.3\%, however, the highest level accuracy is approximately 82.3\%.
\paragraph*{Experiment 3: Image Data} Overall, the accuracy loss is around 40\% which is greater than both previous experiments, but not significant: the lowest level of accuracy is 54.9\%, however, the highest level accuracy is approximately 57.34\%.

Clearly, the analytical algorithm that is used has an influence with trade-offs. For example, the algorithms are optimized for this localized setting as well as the used local processing power that can impact on the accuracy. The correct balance in terms of data transmission, privacy, energy consumption, accuracy and resource cost will need to be identified and our future work will further this area. The results show that the traditional architecture which is based on sending all the data from data source to a single server point has limitations. This is particularly evident when using large data volume with restrictions on time. The experiments show how the data communication over the network can be very expensive while sending large data volume without applying any transformation in advance. Additionally, it shows how effective our hybrid approach is in this regard.

 Based on the result, in text data, there was no significant difference in total processing time among the four architectures. The reason for this might be the data transformation algorithms that are used. However, the idea is not just to reduce the size of data, but also to create meaningful data to get more insights. We have used three types of datasets including numerical, text and image. These datasets are not taken from an industrial application. The reason is that there are no publicly available datasets to conduct our experiments. However, the types of data in the used datasets are quite similar to industrial data in terms of numerical, text and image data. The survey in \cite{perera2014survey} has reviewed over 100 Internet of Things solutions and divided them in the industry marketplace into five classes including smart home, smart wearable, smart environment,smart city and smart enterprise. The datasets that we have used in our experiments fall into some of these classes in terms of data types (numeric, text and image). 

For completeness, while working with the image dataset, one of the difficulties faced was installing imutils library \footnote{https://pypi.python.org/pypi/imutils} and OpenCV library \footnote{https://opencv.org/} for python. Also, we used Linux based system for our experiments because we faced issues with installing libraries/packages and adapting the environments for the experiments while using other operating systems. Additionally, the accuracy of the image dataset might be a bit low, but it is possible to increase the accuracy by using different methods than KNN such as Convolutional Neural Networks (CNN). Similarly, in both datasets including numerical and text, higher accuracy can be obtained either by tuning the machine learning algorithms or by using different algorithms such as CNN. In addition, the data transformation methods can be utilised or different methods can be used to create features with more insights. These issues can be important aspects of data analytics by finding answers regarding how to increase the accuracy, but it is important to use lightweight solutions to avoid having high execution time. However, in this paper, we focused on exploring the most effective fog computing architectures for the IoT. 
 \begin{figure}[!b]
  \centering
   \begin{minipage}[b]{0.48\textwidth}
    \includegraphics[width = 8.6cm, height = 11cm, frame]{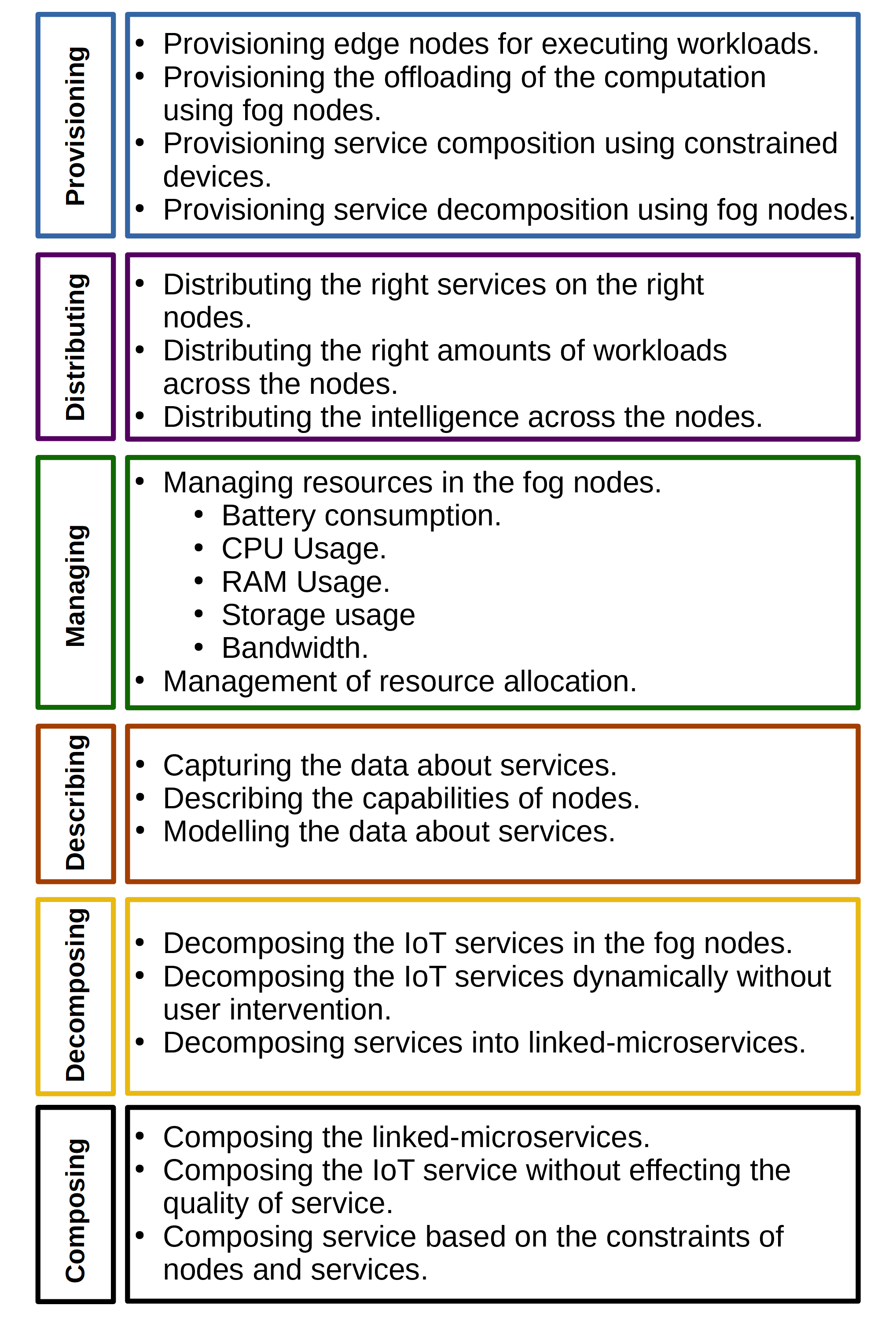}
  \end{minipage}
\caption{Research Challenges}
\label{Research Challenges}
\end{figure}
\section{Requirement Gathering and Research Challenges}
Constraint awareness is an important aspect of the IoT architecture as it will connect a large number of devices with varying computational capabilities, storage, battery power and Internet connectivity. Further, there will be a variety of services with different requirements (e.g. resource requirements, data requirements, latency requirements). The main challenge is to determine which services should be run on which node in a given IoT architecture, by considering both overall efficiency and feasibility. In addition, the management of resources at the edge of a network is crucial for evaluating the potential of fog computing. However, this is challenging in the IoT for a number of reasons. We present below several research challenges, as shown in Figure \ref{Research Challenges} that need to be overcome for practical realisations of fog computing in the IoT domain. Personal data store \cite{Valorising} is an example of where addressing these challenges becomes a necessity. 

\subsection{Provisioning} 
The first challenge is the provisioning of edge nodes for executing workloads that are offloaded by other fog nodes \cite{zhou2017} or from cloud servers \cite{wang2017enorm}. The reduced hardware and processing configurations, heterogeneity of available resources across the range of possible edge nodes and the "lack of standard protocols for initialising services on a potential edge node" \cite{wang2017enorm} add to the complexity. However, harnessing the capabilities of the diverse resources can contribute to extending the boundaries of a cloud system and provide additional revenue models for network providers, by offering incremental data processing as it moves from the source to its destination \cite{zamani2017deadline}. Moreover, matching service execution requirements to the fog nodes' available configurations is also key, necessitating composition or decomposition. For example, consider a service which needs a device configuration of a quad-core processor, 2 GB RAM and 4 GB Storage. Executing this service on most fog nodes is not possible due to their limited processing power. Therefore, there is a need to decompose this service into smaller services, which comes with its own challenges. Similarly, for service composition, it is not possible to provide a composition of two services with high resource demands on fog nodes. 

\subsection{Distribution} 
The second challenge is distributing the workload on the fog nodes. The lack of industry-standard application containers or Virtual Machines for the diverse edge devices makes it difficult to seamlessly distribute the computation across the edge devices.  Moreover, it is not possible to process large workloads on fog nodes as mentioned previously, due to the limited processing power. It is difficult to make a decision about the amount of computation load that can be assigned to a fog node. Moreover, distributing the intelligence across the fog nodes is challenging since most of the neural network, artificial intelligence and machine learning algorithms require high processing power. Current research addresses this issue by implementing different optimization strategies which prioritize different aspects, such as a number of resources without violating application QoS \cite{Mahmud2017Latencyaware}, subjective notions of the value of data to the user to decide the location of data processing \cite{zamani2017deadline}, or prioritising the device's primary function over offloaded workloads \cite{wang2017enorm}. 

\subsection{Resource Management} 
The third challenge is resource management in the fog nodes. Due to their limited computation power, distributing the workloads to edge nodes is challenging because this needs to be done dynamically with the given limited configuration of resources. Therefore, managing resources like battery consumption, CPU usage, RAM usage, storage usage and bandwidth are difficult in an environment that changes dynamically and unexpectedly which makes the process of resource allocation difficult as well. 

\subsection{Describing Nodes} 
The fourth challenge is discovering and describing node capabilities, made especially difficult due to the heterogeneous and volatile nature of the IoT, making it difficult to capture and model data about offered services which have to be done dynamically. While there exist some efforts for a standardised terminology for constrained devices, such as the RFC 7228 \cite{RFC7228}, there is a need to describe their capabilities dynamically which is challenging and impossible to do manually. Analogously, there is a need to capture and model the data about the services (e.g. RAM usage, CPU usage and others) to orchestrate and allocate them to the right nodes.

\subsection{Decomposition and Composition}
The fifth challenge is decomposition; as we have shown in this paper, decomposition plays an important role in increasing the quality of service, but decomposing IoT services dynamically and in an automated way is challenging in fog nodes due to the resource constraints and the dynamicity of IoT. In addition, decomposing the services in linked-microservices is challenging because it is difficult to create services that distributed to different nodes and linked to their linked-partners. Similarly, in the composition process, this will be challenging because identifying linked-microservices which are distributed across the network and then composing them is a sophisticated process in an IoT environment. Additionally, this has to be achieved without affecting the quality of service and data, while using constrained devices.

Most of the proposals in service computing domain usually do not give much attention to the nodes in IoT, focussing instead on the quality of service and service constraints requirements. However, in the case of IoT systems, there will be many constrained devices distributed across the network. This means it is important to decompose a resource-heavy service into smaller micro-linked services which can be handled by constrained devices. Then, distributing these micro-linked services to the nodes based on their capabilities is important. As IoT devices have constraints on resources like RAM, CPU and storage and the services have restrictions on the same resources. Therefore, we need to model the data about services to get the knowledge about their restrictions before distributing them across the nodes. For example, in this paper, we conducted an experiment that decomposes the services to linked-microservices and then we distributed the linked-microservices to fog and cloud based on their capabilities manually.  This is particularly difficult in IoT systems because of the two key characteristics of IoT systems: the heterogeneity and volatility. Therefore, this description of services and nodes need to be modelled autonomously. Possible solutions to node description can be extending the existing Web Service Description Language (WSDL), creating node description language etc. 
Additionally, a possible solution to the distribution of services can be artificial intelligence (AI) planning techniques.

\section{Conclusion}
This paper presents an efficient approach where the raw data is preprocessed in the fog node before being transmitted to the cloud to minimise the data communication over the network. Three types of datasets are used for the experiments including numerical data, text data and image data. Furthermore, we conducted 4 experiments, including 4 architectures namely cloud, fog, hybrid and fog + cloud for each dataset to explore which one is the most effective for the Internet of Things. The results show that the hybrid approach is efficient in terms of minimizing the cost of data communication over the network while maintaining the accuracy. However, fog + cloud approach could be useful to employ in situations where fog device have limited processing capability and cannot perform all the required processing. In such situations, fog + cloud  approach would perform better than cloud only approach. In addition, we used the WISDM dataset \cite{kwapisz2011activity}, 20 Newsgroups dataset \cite{Lang95} and Kaggle dogs vs cats dataset \cite{dogsvscats|kaggle_2013} to validate our architecture. 

Future research will involve distributing services to nodes dynamically in an optimum way. Moreover, research questions that could be asked include which service(s) should be run on which node by being aware of constraints. In addition, using different algorithms to extract different features for greater data transformation. Additionally, further minimizing data communication over the network while maintaining the accuracy. Furthermore, consideration of the positive effect on privacy and evaluation of energy consumption will be elements to consider in future work.
\section*{Acknowledgment}
 Badraddin Alturki's research is funded by the Saudi Arabian Cultural bureau in London and his scholarship is granted by King Abdul Aziz University. Dr De's research is  funded by the TagItSmart! collaborative project supported by the European Horizon 2020 programme, contract number: 688061. Dr Perera's work is supported by EPSRC PETRAS 2 (EP/S035362/1).
\bibliographystyle{IEEEtran}
\bibliography{scholar}
\vskip -2\baselineskip plus -1fil

\begin{IEEEbiography}[{\includegraphics[width=1in,height=1.25in,clip,keepaspectratio]{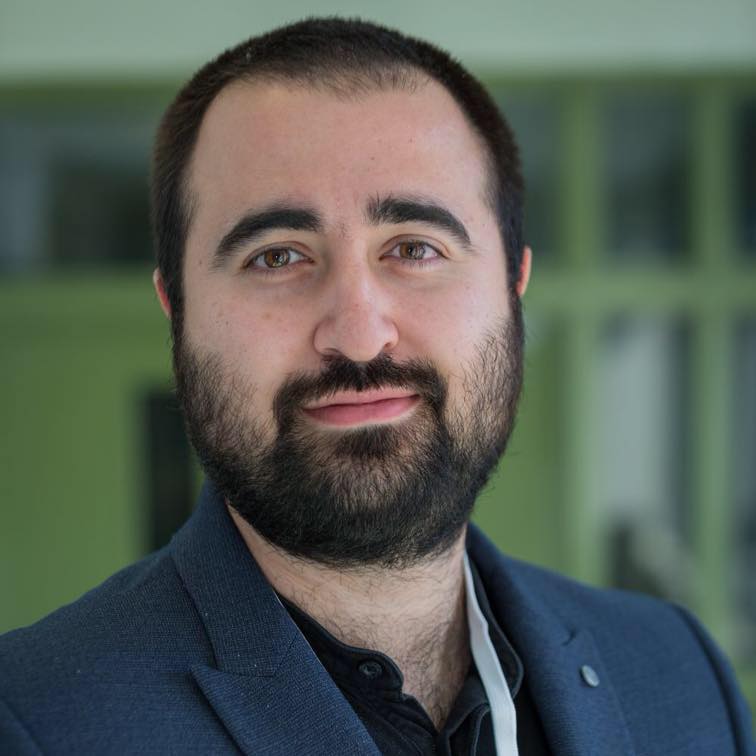}}]{Badraddin Alturki}received the Masters degree in Software Engineering from
the University of Leicester, Leicester, U.K. in 2015. He is PhD Candidate in Computer Science at the
University of Leicester. His research work focus on distributing services on nodes in fog computing for the Internet of Things.
He is a member in both ACM and IEEE

\end{IEEEbiography}
\begin{IEEEbiography}[{\includegraphics[width=1in,height=1.25in,clip,keepaspectratio]{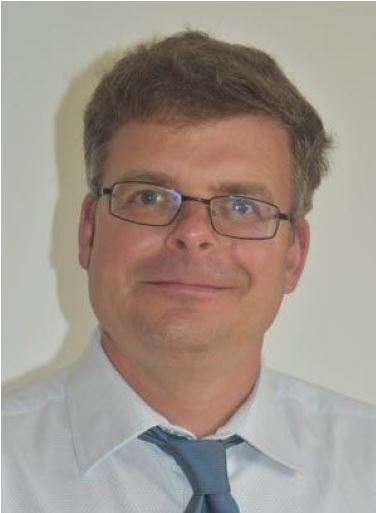}}]{Stephan Reiff-Marganiec}is a Senior Lecturer
(Associate Professor) in Computer Science at the
University of Leicester. He has worked in the computer
industry in Germany and Luxembourg and
held research positions at the University of Glasgow
(while simultaneously reading for a PhD) and the
University of Stirling. He co-chaired the 8th and
10th International Conference on Feature Interactions
in Telecommunications and Software Systems
an was co-Chair of three instances of YR-SOC.
Stephan lead work packages in the EU funded
projects Leg2Net, Sensoria and inContext focusing on automatic service
adaption, context aware service selection, workflows and rule based service
composition. Stephan is co-editor of the Handbook of Research on Service-
Oriented Systems and Non-Functional Properties and has published in excess
of 50 papers in international conferences and journals as well as having served
on a large number of programme committees. Stephan was appointed Guest
Professor at the China University of Petroleum and was visiting Professor at
Lamsade at the University of Dauphine, Paris. He was elected Fellow of the
BCS (FBCS) in 2009 and is a member in both ACM and IEEE.

\end{IEEEbiography}
\begin{IEEEbiography}[{\includegraphics[width=1in,height=1.25in,clip,keepaspectratio]{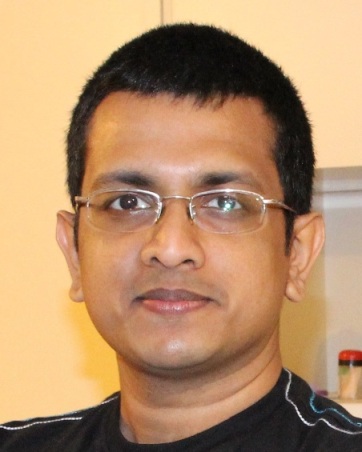}}]{Charith Perera} is a Lecturer (Assistant Professor) in Computing Science at Cardiff University, United Kingdom. Dr. Perera received the B.Sc. (Hons) degree in computer science from Staffordshire University, Stoke-on-Trent, UK. in 2009 and the Master of Business Administration (MBA) from the University of Wales, Cardiff, UK. in 2012. He received the Ph.D. degree in computer science with The Australian National University, Canberra, Australia. He completed his post-doctoral at Newcastle University, UK and Open University, UK. Previously, he was with the Information Engineering Laboratory, ICT Centre, CSIRO, Australia. His research interests include Internet of Things, Sensing as a Service, infrastructure and architectures, privacy and security. Dr. Perera is a member of the ACM and IEEE. For more details: charithperera.net

\end{IEEEbiography}
\begin{IEEEbiography}[{\includegraphics[width=1in,height=1.25in,clip,keepaspectratio]{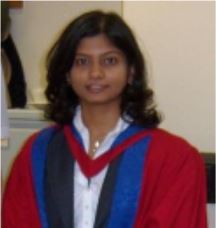}}]{Suparna De}is a Senior Research Fellow at the Institute for Communication Systems (ICS), University of Surrey, UK. She obtained her Ph.D. and MSc. (with distinction) degrees in Electronic Engineering from the University of Surrey in 2009 and 2005, respectively. She has been leading technical work areas related to various aspects of service provisioning and data analysis in the Internet of Things domain in several EU projects such as TagItSmart, iKaaS, IoT.est, iCore and IoT-A. Her research has been supported by grants from the EC H2020 and FP7 programs and through DTI, UK-funded programs. Her research interests include large-scale data analytics and latent pattern discovery in Web of Things scenarios, semantic association analysis and knowledge engineering methods. She is a member of both the IEEE and ACM.

\end{IEEEbiography}

\end{document}